\documentclass[aps,prx,reprint,superscriptaddress,nofootinbib]{revtex4-2}
\usepackage{graphicx,amsmath,amssymb,bm,xcolor,booktabs}
\usepackage{tikz}
\usepackage{algpseudocode}
\newcounter{algorithm}
\newenvironment{algbox}[1]{\begin{figure}[!tbp]\refstepcounter{algorithm}\hrule\vspace{3pt}%
  \noindent\small\textbf{Algorithm~\thealgorithm}\enspace #1\par\vspace{3pt}\hrule\vspace{3pt}\footnotesize}%
  {\vspace{2pt}\hrule\end{figure}}
\usetikzlibrary{positioning,arrows.meta,calc,fit,backgrounds,decorations.pathreplacing}
\usepackage[colorlinks=true,linkcolor=blue!50!black,citecolor=blue!50!black,urlcolor=blue!50!black]{hyperref}

\graphicspath{{./}{figures/}}
\newcommand{\kT}{k_BT}
\newcommand{\pe}{p_{\mathrm{L}}}

\begin{document}

\title{Reinforcement Learning Control of Information Erasure in Quantum Error Correction}

\author{Ramon Jose C. Bagunu}
\affiliation{Department of Physical Sciences and Mathematics, College of Arts and Sciences,\\ University of the Philippines Manila, Manila 1000, Philippines}
\email{rcbagunu@up.edu.ph}

\date{\today}

\begin{abstract}
Every round of quantum error correction (QEC) erases information. Standard protocols erase every round, and by Landauer's principle every erasure dissipates heat. In this paper we ask how much of that erasure protection actually needs. The decisions made in each round, whether to check or skip, keep or erase the syndrome, and correct now or in software, are cast as the control of entropy flow, with a cost equal to the logical error plus a price on the heat released. For a three-qubit repetition code under a noise model fitted to IBM superconducting processors, an exact Bayes-optimal solver and a reinforcement-learning agent trained with proximal policy optimization (PPO) arrive at the same answer. The optimal controller checks once, just before readout, and when heat is priced it undoes the record rather than erasing it. This single late check lowers the logical error by 36--44\% relative to every fixed schedule we tested, from 4 to 10 rounds, where exact search becomes infeasible. Gradient-based learners stall at a local optimum at ``never check'', which a curriculum in the heat price overcomes. In pre-registered, drift-controlled experiments on two IBM processors, frequent checking is confirmed to cost logical error (up to $12\sigma$), checks do not damage the data, and one syndrome ancilla raises a false alarm about 3.5\% of the time during checks. In an exploratory comparison, the late check beats passive majority vote on healthy qubits ($0.0013\pm0.0003$, $4.0\sigma$), but not in two pre-registered runs in which a data qubit had degraded. At the scale of current devices, much of the heat paid for active error correction therefore appears to buy no additional protection.
\end{abstract}

\keywords{quantum error correction, information erasure, Landauer principle, reinforcement learning, quantum control, superconducting qubits}

\maketitle

%% ===================================================================== INTRODUCTION
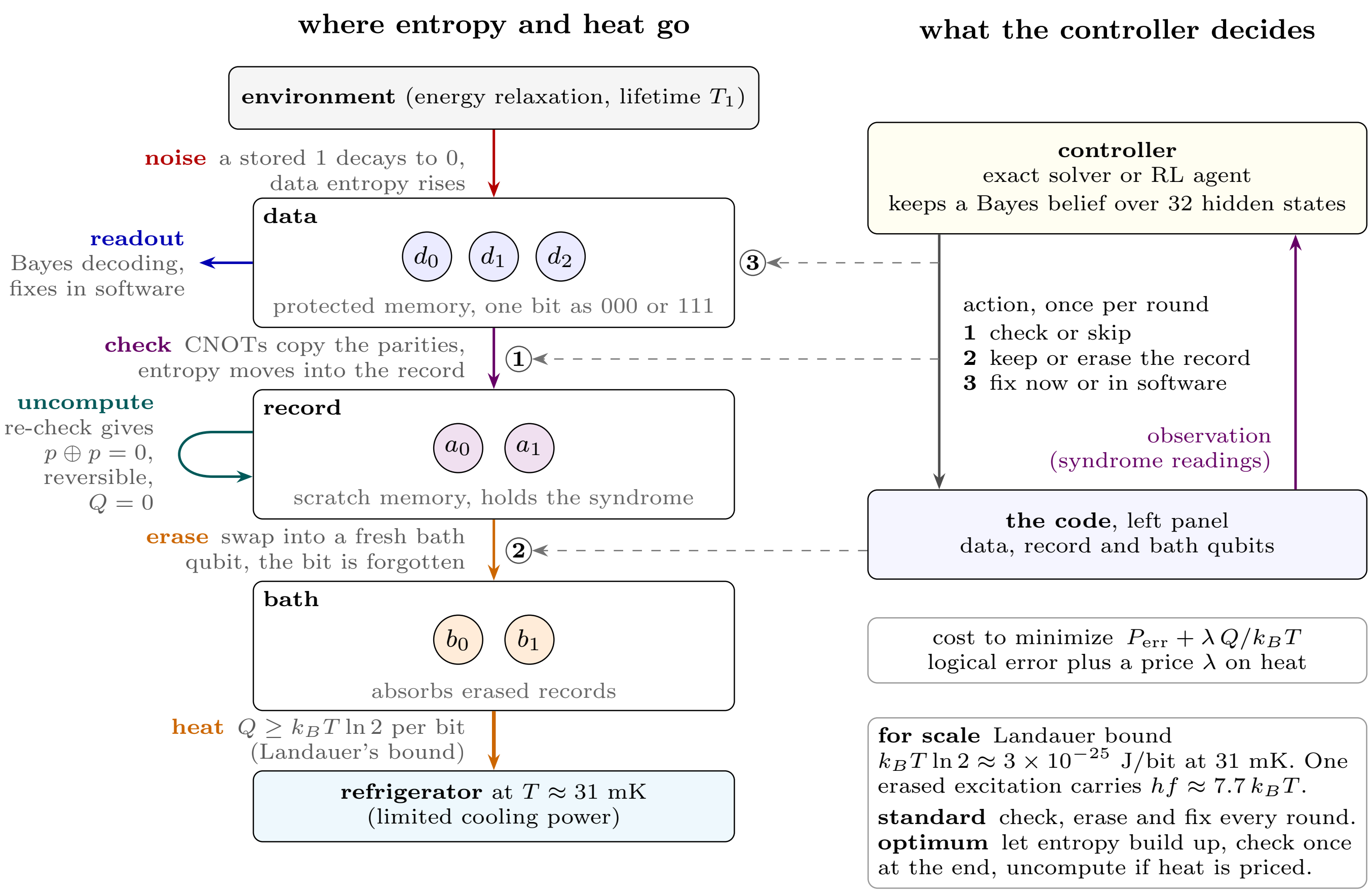
\begin{figure*}[!t]
\centering
\resizebox{\textwidth}{!}{%
\begin{tikzpicture}[font=\footnotesize, >={Stealth[length=1.8mm]},
  dq/.style={circle,draw,fill=blue!8,minimum size=5.5mm,inner sep=0},
  aq/.style={circle,draw,fill=violet!12,minimum size=5.5mm,inner sep=0},
  bq/.style={circle,draw,fill=orange!15,minimum size=5.5mm,inner sep=0},
  reg/.style={draw,rounded corners=3pt,minimum width=5.4cm,minimum height=1.45cm,inner sep=4pt},
  ttl/.style={anchor=north west,font=\scriptsize\bfseries},
  mean/.style={font=\scriptsize,text=black!60},
  lab/.style={align=right,font=\scriptsize,anchor=east},
  dec/.style={circle,draw=black!70,fill=white,inner sep=0.6pt,font=\scriptsize\bfseries},
  ctl/.style={->,dashed,black!65},
  flow/.style={->,thick}]
% ---------------- registers
\node[reg,fill=gray!8,minimum height=7mm,font=\scriptsize] (env) at (0,0) {\textbf{environment} (energy relaxation, lifetime $T_1$)};
\node[reg] (data) at (0,-1.85) {};
\node[ttl] at (data.north west) {data};
\node[dq] at (-0.75,-1.78) {$d_0$}; \node[dq] at (0,-1.78) {$d_1$}; \node[dq] at (0.75,-1.78) {$d_2$};
\node[mean] at (0,-2.35) {protected memory, one bit as 000 or 111};
\node[reg] (rec) at (0,-4.0) {};
\node[ttl] at (rec.north west) {record};
\node[aq] at (-0.4,-3.93) {$a_0$}; \node[aq] at (0.4,-3.93) {$a_1$};
\node[mean] at (0,-4.5) {scratch memory, holds the syndrome};
\node[reg] (bath) at (0,-6.15) {};
\node[ttl] at (bath.north west) {bath};
\node[bq] at (-0.4,-6.08) {$b_0$}; \node[bq] at (0.4,-6.08) {$b_1$};
\node[mean] at (0,-6.65) {absorbs erased records};
\node[reg,fill=cyan!6,minimum height=7mm,font=\scriptsize,align=center] (fr) at (0,-7.95) {\textbf{refrigerator} at $T\approx31$ mK\\(limited cooling power)};
% ---------------- physical flows (labels on the left)
\draw[flow,red!70!black] (env) -- (data);
\node[lab,text=red!70!black] at (-0.2,-0.83) {\textbf{noise}\enspace \textcolor{black!65}{a stored 1 decays to 0,}\\\textcolor{black!65}{data entropy rises}};
\draw[flow,violet!80!black] (data) -- (rec);
\node[lab,text=violet!80!black] at (-0.2,-2.93) {\textbf{check}\enspace \textcolor{black!65}{CNOTs copy the parities,}\\\textcolor{black!65}{entropy moves into the record}};
\draw[flow,orange!80!black] (rec) -- (bath);
\node[lab,text=orange!80!black] at (-0.2,-5.08) {\textbf{erase}\enspace \textcolor{black!65}{swap into a fresh bath}\\\textcolor{black!65}{qubit, the bit is forgotten}};
\draw[flow,orange!80!black,very thick] (bath) -- (fr);
\node[lab,text=orange!80!black] at (-0.2,-7.22) {\textbf{heat}\enspace \textcolor{black!65}{$Q\ge k_BT\ln2$ per bit}\\\textcolor{black!65}{(Landauer's bound)}};
% uncompute loop on the left of the record
\draw[flow,teal!70!black] ([yshift=2.5mm]rec.west) -- ++(-0.45,0) .. controls +(-0.5,0) and +(-0.5,0) .. ++(0,-0.5) -- ([yshift=-2.5mm]rec.west);
\node[lab,text=teal!70!black] at (-3.7,-4.0) {\textbf{uncompute}\\\textcolor{black!65}{re-check gives}\\\textcolor{black!65}{$p\oplus p=0$,}\\\textcolor{black!65}{reversible,}\\\textcolor{black!65}{$Q=0$}};
% readout on the left of the data
\draw[flow,blue!70!black] (data.west) -- ++(-0.6,0);
\node[lab,text=blue!70!black] at (-3.35,-1.85) {\textbf{readout}\\\textcolor{black!65}{Bayes decoding,}\\\textcolor{black!65}{fixes in software}};
% ---------------- panel labels
\node[anchor=south,font=\small\bfseries] at (0,0.55) {where entropy and heat go};
\node[anchor=south,font=\small\bfseries] at (7.0,0.55) {what the controller decides};
% decision markers on the flows (keyed to panel b)
\node[dec] (m1) at (0.28,-2.93) {1};
\node[dec] (m2) at (0.28,-5.08) {2};
\node[dec] (m3) at ([xshift=2mm]data.east) {3};
% ---------------- panel (b): agent-environment loop
\node[draw,rounded corners=3pt,fill=yellow!6,align=center,minimum width=5.6cm,minimum height=1.25cm,font=\scriptsize] (agent) at (7.0,-0.9) {\textbf{controller}\\exact solver or RL agent\\[1pt] keeps a Bayes belief over 32 hidden states};
\node[draw,rounded corners=3pt,fill=blue!4,align=center,minimum width=5.6cm,minimum height=1.0cm,font=\scriptsize] (code) at (7.0,-4.9) {\textbf{the code}, left panel\\data, record and bath qubits};
\draw[->,thick,violet!80!black] ([xshift=2.0cm]code.north) -- ([xshift=2.0cm]agent.south);
\node[anchor=east,align=right,font=\scriptsize,text=violet!80!black] at (8.85,-3.95) {observation\\(syndrome readings)};
\draw[->,thick,black!70] ([xshift=-2.0cm]agent.south) -- ([xshift=-2.0cm]code.north);
% actions enter panel (a) at the numbered points
\draw[->,dashed,black!55] (5.0,-1.85) -- (m3);
\draw[->,dashed,black!55] (5.0,-2.93) -- (m1);
\draw[->,dashed,black!55] (code.west |- m2) -- (m2);
\node[anchor=west,align=left,font=\scriptsize] at (5.15,-2.75) {action, once per round\\[1pt]
\textbf{1}\enspace check or skip\\
\textbf{2}\enspace keep or erase the record\\
\textbf{3}\enspace fix now or in software};
\node[draw=black!40,rounded corners=3pt,align=center,minimum width=5.6cm,font=\scriptsize] (cost) at (7.0,-6.2) {cost to minimize\enspace $P_{\rm err}+\lambda\,Q/k_BT$\\logical error plus a price $\lambda$ on heat};
\node[draw=black!40,rounded corners=3pt,align=flush left,text width=5.38cm,font=\scriptsize,anchor=north] at (7.0,-6.95) {%
\textbf{for scale}\enspace Landauer bound $k_BT\ln2\approx3\times10^{-25}$ J/bit at 31 mK. One erased excitation carries $hf\approx7.7\,k_BT$.\\[2pt]
\textbf{standard}\enspace check, erase and fix every round.\\
\textbf{optimum}\enspace let entropy build up, check once at the end, uncompute if heat is priced.};
\end{tikzpicture}}
\caption{Entropy and heat flow in one round, and where the controller acts. Left, decay raises the entropy of the data, and a check moves it into the record. The record then leaves either by erasure into the bath, which releases heat $Q\ge\kT\ln2$ per bit, or by uncomputation ($p\oplus p=0$), which releases none. Right, the controller sees only the syndrome readings and makes three decisions per round, which act at the numbered points on the left. The whole code on the left is its RL environment.}
\label{fig:entropy}
\end{figure*}

\section{Introduction}

Quantum error correction protects information by repeatedly discarding information. Its task is to keep a fragile quantum state intact while every physical qubit holding it slowly loses its information to the environment. The standard approach, in its simplest form, is redundancy together with measurement. A logical bit is spread over several physical qubits, and the parities between them, the syndrome, are measured repeatedly so that errors can be located and undone~\cite{Shor1995,Steane1996,NielsenChuang2010,Terhal2015}. The syndrome is then written into ancilla qubits, and before those ancillas can be used again, they must be returned to a known state.

That reset is information erasure in Landauer's sense. A memory holding one bit of unknown value is forced into a fixed value, which is logically irreversible and dissipates at least $\kT\ln 2$ of heat~\cite{Landauer1961,Berut2012,Parrondo2015,ReebWolf2014}. This should not be confused with erasure \emph{errors}, the loss of a qubit, which the QEC literature also calls ``erasure''. In superconducting processors the heat of each reset is released at the coldest stage of the refrigerator, and the total grows with the number of physical qubits and the number of rounds~\cite{Auffeves2022,FellousAsiani2023,Bilokur2024}. Recent work has bounded the work needed to keep a syndrome memory reusable~\cite{Czartowski2026}, cast QEC as a heat engine~\cite{Landi2020QECengine}, and reduced the heat of each syndrome erasure on IBM hardware by recycling an out-of-equilibrium bath~\cite{IshidaHasegawa2026}. Together these works suggest a useful way to view a QEC cycle. Errors raise the entropy of the data, checks move that entropy into the ancillas, and erasures discard it into the bath, so that the cycle works as a refrigerator for the logical information.

A complementary question has received much less attention. How hard does this refrigerator need to run, and how much of its erasure is needed at all? The standard cycle checks and resets every round~\cite{Fowler2012}, and repetition- and surface-code experiments on current hardware run it this way~\cite{Kelly2015,GoogleRepetition2021,GoogleThreshold2025}. There is growing evidence that checking less often can help when the checks themselves are noisy~\cite{AbuNada2017,MartonAsboth2025,Haug2026,Berthusen2025,SohnOh2026}. These studies, however, optimize the measurement schedule for logical error alone. They do not price heat, and they do not let the controller decide whether to keep or erase the record once it has been written.

To make the distinction concrete, consider two controllers side by side. The standard controller checks every round, erases the record into a bath every round, and applies a correction immediately whenever the syndrome fires. A free controller makes each of those three choices separately in every round, using everything it has observed so far. It may skip a round, keep a record on the ancillas and undo it later with a second check, or postpone every correction to the final readout. The standard controller is one particular point in the free controller's space of options. The question is whether it is a good one, and if not, what the best point looks like when both protection and heat are counted. Figure~\ref{fig:entropy} shows the whole setup, with the flow of entropy and heat in the code on the left and the controller that makes these choices on the right.

The answer, for a small code on current IBM hardware, is simpler than one might expect. The optimal controller does not use the free controller's flexibility to check cleverly. Instead, it almost never checks. When heat is free it checks exactly once, just before readout, and when heat costs something it undoes its record rather than erasing it. On hardware, the costs of checking often are confirmed. The advantage of the single late check, however, survives only on healthy qubits, and tracing why led us to a property of the syndrome measurement itself.

We obtain this answer for a three-qubit repetition code under a noise model fitted to IBM superconducting processors. The per-round decisions are treated as a control problem whose cost is the logical error plus a price $\lambda$ on the heat released. We solve it in two independent ways, by exact Bayes-optimal search over the full decision tree and by reinforcement learning (RL)~\cite{SuttonBarto2018,Schulman2017}. RL has been applied to quantum feedback and state preparation~\cite{Fosel2018,Bukov2018}, to error correction and code discovery~\cite{Sivak2023,Nautrup2019} and to decoding~\cite{Sweke2021}, and machine learning more broadly has produced decoders that outperform hand-designed ones on hardware data~\cite{Bausch2024}. RL has also been used to optimize quantum thermal machines and to reduce entropy production~\cite{ErdmanNoe2022,Sgroi2021}. Most recently, RL has been used to steer the control parameters of a running surface-code experiment, using error-detection events as the learning signal, so that the code stays calibrated as the device drifts~\cite{Sivak2025RLQEC}. That work tunes the physical operations of a fixed QEC cycle. Here we leave the operations fixed and control the cycle itself, deciding when to check, whether to keep or erase each record, and when to correct, with the heat released as part of the cost. The two kinds of control act at different levels and could be combined. Here the exact solver lets us check the learner where the answer is known before relying on it where brute force is out of reach. We then test the resulting controllers on two IBM processors in pre-registered, drift-controlled experiments.

In this paper we make four contributions.
\begin{enumerate}
\item We formulate error correction as the control of entropy flow out of the data, with an explicit heat cost and the option to keep a record and later undo it at no heat cost. The formulation is defined for any stabilizer code, and we solve it exactly for a three-qubit repetition code under a hardware-fitted noise model.
\item We show that the optimal controller erases almost nothing, and that a PPO agent recovers it. The resulting single-late-check timetable outperforms every fixed schedule we tested, standard, periodic or passive, by 36--44\% in simulation from 4 to 10 rounds, where brute-force exact search is no longer practical.
\item We explain where learning gets stuck, and why. The problem has a local optimum at ``never check'' that is two actions wide, which exploration only partly escapes and a curriculum in the heat price overcomes. A deep Q-network (DQN), by contrast, fails at high heat prices through value overestimation.
\item We report pre-registered, drift-controlled experiments on two IBM processors that locate where the simulated advantage is lost. It is lost not in damage to the data but in syndromes that are less reliable during checks than in readout, and in a sensitivity to degrading data qubits that we report as an open problem.
\end{enumerate}

The three-qubit code is chosen so that every claim about the optimum can be checked exactly and every deviation on hardware can be traced. The formulation, the use of the exact optimum as a benchmark for the learner, and the in-job calibration are built to be reused when the code grows, the errors include dephasing, and exact search is no longer available.

The rest of the paper is organized as follows. Section~\ref{sec:background} summarizes the physical and learning concepts the paper relies on. Section~\ref{sec:setting} describes the code, the actions, the noise model and the heat accounting. Section~\ref{sec:methods} describes the exact solver, the learning agents and the hardware protocol. Sections~\ref{sec:optimum} and~\ref{sec:learning} present the optimal controller and what the learners do. Section~\ref{sec:hardware} presents the hardware experiments, and Sec.~\ref{sec:discussion} discusses implications, limitations and practical recommendations.

%% ===================================================================== BACKGROUND

\section{Background}
\label{sec:background}

This work draws on quantum error correction and on reinforcement learning. We summarize here the ideas from each that the rest of the paper relies on, starting with the physical system and the heat cost of resetting it and then turning to the learning framework used to control it.

\subsection{Qubits, syndromes and the cost of erasure}

\emph{Qubits and decay.} A superconducting qubit is a small electrical circuit whose two lowest energy levels, $|0\rangle$ and $|1\rangle$, store one bit. The excited level $|1\rangle$ is not stable and it decays to $|0\rangle$ at random, with a characteristic time $T_1$ of a few hundred microseconds on current IBM devices. The decay is one-way, since a qubit at a few tens of millikelvin almost never jumps up spontaneously. For a bit stored in such qubits, the typical error is therefore a $1$ that becomes a $0$, and the probability of such an error grows with the time the qubit waits.

\emph{Readout.} A qubit is read out by probing its circuit with a microwave pulse, which takes about $2\,\mu$s and is itself imperfect. A $0$ can be read as $1$ and, more often, a $1$ as $0$, because the qubit can decay during the readout. Each qubit has its own pair of readout errors, $r_{01}$ and $r_{10}$, of the order of $0.1$--$3\%$.

\begin{table*}[!t]
\caption{Heat for one error pattern. Logical $1$ is stored and $d_1$ decays before round 1 of 2, so a check reads $11$ until $d_1$ is fixed. Each round entry gives the reading and what the controller does with it. Every $1$ erased into the bath carries $x\kT$ of heat, and anything left on the ancillas is erased at the end.}
\label{tab:example}
\begin{ruledtabular}
\begin{tabular}{lccccc}
controller & round 1 & round 2 & end & 1s erased & heat \\
\hline
standard (fix now, erase) & $11$, fix, erase & $00$, erase & -- & 2 & $2x\kT$ \\
erase, fix in software & $11$, erase & $11$, erase & -- & 4 & $4x\kT$ \\
keep, then check again & $11$, keep & $00$ (undone) & -- & 0 & $0$ \\
check once at the end & skip & $11$, keep & erase & 2 & $2x\kT$ \\
\end{tabular}
\end{ruledtabular}
\end{table*}

\emph{Repetition code and syndrome.} The simplest protection against bit flips is redundancy. A logical $0$ is stored as $000$ and a logical $1$ as $111$, and at the end the majority of the three readings is taken. Here a single flipped qubit is outvoted. Error correction goes further by measuring, during the computation, the parities $d_0\oplus d_1$ and $d_1\oplus d_2$ without measuring the data themselves. These two parities are the \emph{syndrome}. A syndrome of $00$ means the three data agree, while $10$, $11$ and $01$ point at a flip of $d_0$, $d_1$ or $d_2$. The parities are copied onto two extra \emph{ancilla} qubits by CNOT gates, and the ancillas are then read out. Only three gates are needed, and because the qubits here only ever hold $0$ or $1$, each acts as a simple rule on bits. The $X$ gate flips a bit, $b\to1-b$, and the CNOT flips its target when its control is $1$, $(c,t)\to(c,\,t\oplus c)$. Here an ancilla starting at $0$ ends up holding the parity of the data qubits that control it. The SWAP exchanges two bits, $(a,b)\to(b,a)$, and is built from three CNOTs. Every gate therefore permutes the bit strings of the register, which is why the state of the code can be tracked as a probability distribution over bit strings (Sec.~\ref{sec:setting}). Measuring an ancilla in the middle of a computation, while the data qubits wait, is called a \emph{mid-circuit measurement}. It is noisier than a final readout, because the data keep decaying during it and because the measurement can disturb neighboring qubits~\cite{Rudinger2022,Hothem2025}.

\emph{Landauer erasure and heat.} Once an ancilla holds a syndrome bit, it must be returned to $0$ before it can be used again. If the controller no longer knows what the ancilla holds, this reset discards one bit of information, and Landauer's principle states that discarding information dissipates at least $\kT\ln2$ of heat into the environment~\cite{Landauer1961,Berut2012}. On hardware the ancilla's state is swapped into a separate bath qubit, whose excitation then leaks into the refrigerator. We measure that heat in units of $\kT$, where $T\approx31$~mK is the refrigerator temperature, so that the Landauer bound is $\ln2\approx0.69$ in these units. However, a reset is not the only way to clear an ancilla. If the controller checks the same parity twice and it has not changed, the second CNOT pair returns the ancilla to $0$. This \emph{uncomputation} is reversible, nothing is forgotten, and Landauer's bound does not apply to it. A controller therefore has two ways to clear the entropy it has moved into an ancilla. It can discard it into the bath as heat, or return it reversibly when the record is no longer needed (Fig.~\ref{fig:entropy}).

\emph{An example.} Table~\ref{tab:example} follows one error pattern through four controllers that differ in whether they check, keep or erase the record, and correct immediately or in software at the end. Logical $1$ is stored, $d_1$ decays before the first of two rounds, and nothing else goes wrong. A check then reads $11$, which blames $d_1$. Each $1$ swapped into the bath carries one excitation of energy $hf$, which is $x\kT$ of heat with $x=hf/\kT\approx7.74$ for a $5$~GHz qubit at $31$~mK (Sec.~\ref{sec:setting}), and a $0$ carries none. The standard protocol pays for one record. Correcting in software while still erasing every round pays twice, because the uncorrected error makes the syndrome fire again in round 2. Then, keeping the record pays nothing because the second check returns the ancillas to $00$ ($p\oplus p=0$). In this example all four controllers decode the logical $1$ correctly. The heat reported in this paper is the average of such counts over all error patterns, weighted by their probabilities under the noise model.

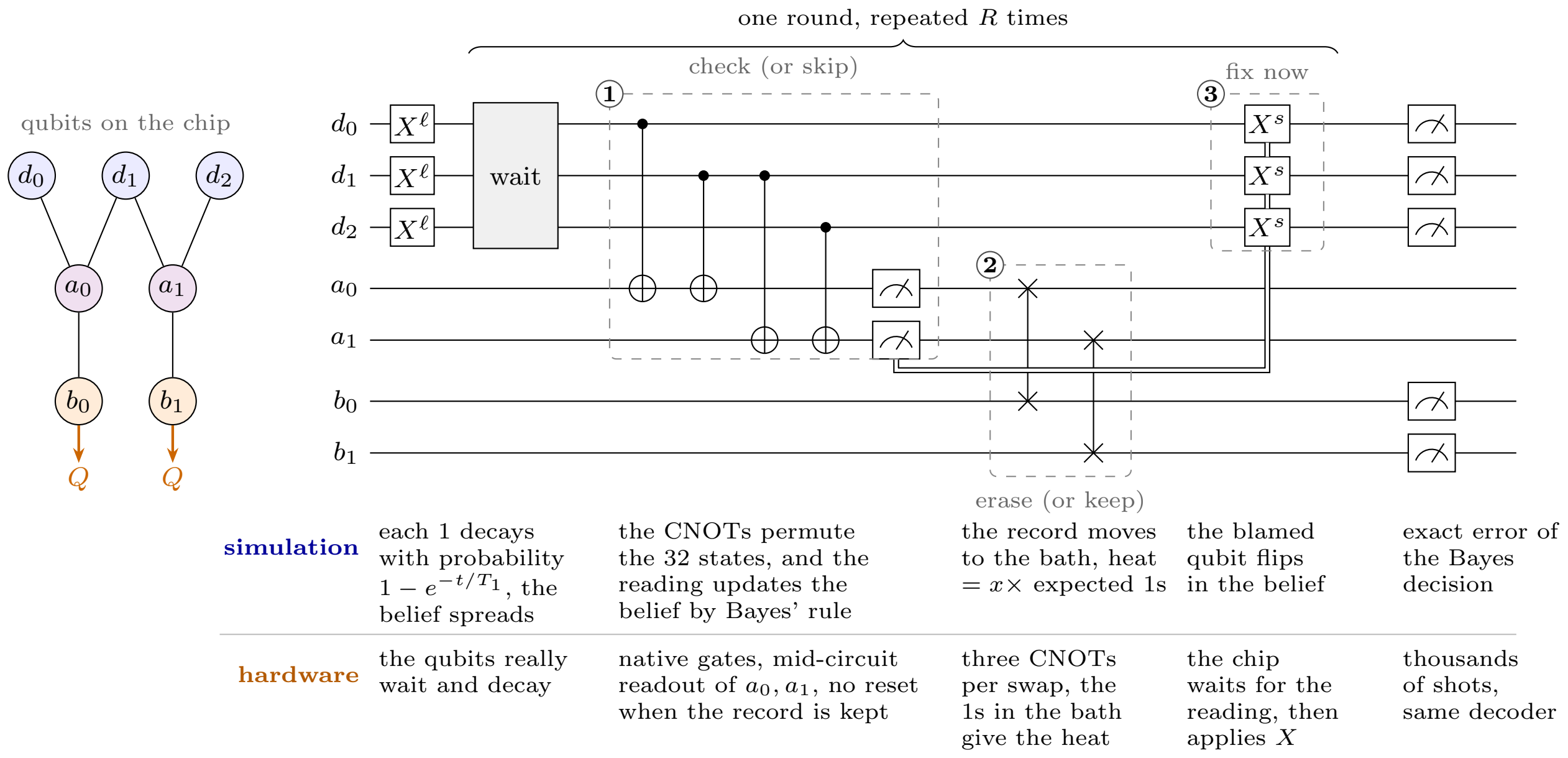
\begin{figure*}[!t]
\centering
\begin{tikzpicture}[font=\footnotesize, >={Stealth[length=1.6mm]},
  dq/.style={circle,draw,fill=blue!8,minimum size=5mm,inner sep=0},
  aq/.style={circle,draw,fill=violet!12,minimum size=5mm,inner sep=0},
  bq/.style={circle,draw,fill=orange!15,minimum size=5mm,inner sep=0},
  gt/.style={draw,fill=white,minimum size=4mm,inner sep=1pt},
  dec/.style={circle,draw=black!70,fill=white,inner sep=0.6pt,font=\scriptsize\bfseries},
  grp/.style={draw=black!45,dashed,rounded corners=2pt},
  note/.style={font=\scriptsize,align=left,anchor=north west,inner sep=1pt,execute at begin node={\hyphenpenalty=10000\exhyphenpenalty=10000}}]
% ---------- layout inset ----------
\begin{scope}[shift={(-3.6,0)}]
\node[dq] (L0) at (0,-0.55) {$d_0$}; \node[dq] (L1) at (1.0,-0.55) {$d_1$}; \node[dq] (L2) at (2.0,-0.55) {$d_2$};
\node[aq] (A0) at (0.5,-1.75) {$a_0$}; \node[aq] (A1) at (1.5,-1.75) {$a_1$};
\node[bq] (B0) at (0.5,-2.95) {$b_0$}; \node[bq] (B1) at (1.5,-2.95) {$b_1$};
\foreach \u/\v in {L0/A0,L1/A0,L1/A1,L2/A1,A0/B0,A1/B1} \draw (\u)--(\v);
\draw[->,orange!80!black,thick] (B0.south) -- ++(0,-0.4) node[below,inner sep=1pt]{$Q$};
\draw[->,orange!80!black,thick] (B1.south) -- ++(0,-0.4) node[below,inner sep=1pt]{$Q$};
\node[font=\scriptsize,text=black!60] at (1.0,0.0) {qubits on the chip};
\end{scope}
% ---------- wires ----------
\def\yA{0}\def\yB{-0.55}\def\yC{-1.1}\def\yD{-1.75}\def\yE{-2.3}\def\yF{-2.95}\def\yG{-3.5}
\foreach \y/\l in {\yA/d_0,\yB/d_1,\yC/d_2,\yD/a_0,\yE/a_1,\yF/b_0,\yG/b_1}{
  \draw (0,\y) -- (12.2,\y); \node[anchor=east] at (0,\y) {$\l$};}
% prepare
\foreach \y in {\yA,\yB,\yC} \node[gt] at (0.45,\y) {$X^{\ell}$};
% idle
\node[draw,fill=black!6,minimum width=0.9cm,minimum height=1.55cm] at (1.55,-0.55) {wait};
% check
\foreach \x/\c/\t in {2.9/\yA/\yD,3.55/\yB/\yD,4.2/\yB/\yE,4.85/\yC/\yE}{
  \draw (\x,\c) -- (\x,\t-0.14); \fill (\x,\c) circle (1.6pt);
  \draw (\x,\t) circle (0.14);}
\foreach \y in {\yD,\yE}{
  \node[gt,minimum width=5mm] (M\y) at (5.6,\y) {}; \draw ($(5.6,\y)+(-0.17,-0.08)$) arc (160:20:0.18); \draw (5.6,\y-0.1) -- ++(0.13,0.2);}
% erase (swaps)
\foreach \x/\u/\v in {7.0/\yD/\yF,7.7/\yE/\yG}{
  \draw (\x,\u) -- (\x,\v);
  \foreach \y in {\u,\v} {\draw (\x-0.1,\y-0.1)--(\x+0.1,\y+0.1); \draw (\x-0.1,\y+0.1)--(\x+0.1,\y-0.1);}}
% fix (feedforward)
\draw[double,double distance=1pt] (5.6,\yE-0.2) -- (5.6,-2.62) -- (9.55,-2.62) -- (9.55,\yC-0.2);
\foreach \y in {\yA,\yB,\yC} \node[gt] at (9.55,\y) {$X^{s}$};
\draw[double,double distance=1pt] (9.55,\yC+0.2) -- (9.55,\yB-0.2) (9.55,\yB+0.2) -- (9.55,\yA-0.2);
% final readout
\foreach \y in {\yA,\yB,\yC,\yF,\yG}{
  \node[gt,minimum width=5mm] at (11.3,\y) {}; \draw ($(11.3,\y)+(-0.17,-0.08)$) arc (160:20:0.18); \draw (11.3,\y-0.1) -- ++(0.13,0.2);}
% groups
\draw[grp] (2.55,0.32) rectangle (6.05,-2.5);  \node[dec] at (2.55,0.32) {1};
\draw[grp] (6.6,-1.5) rectangle (8.1,-3.75);   \node[dec] at (6.6,-1.5) {2};
\draw[grp] (8.95,0.32) rectangle (10.15,-1.35); \node[dec] at (8.95,0.32) {3};
\node[font=\scriptsize,text=black!60,anchor=south] at (4.3,0.36) {check (or skip)};
\node[font=\scriptsize,text=black!60,anchor=north] at (7.35,-3.78) {erase (or keep)};
\node[font=\scriptsize,text=black!60,anchor=south] at (9.55,0.36) {fix now};
% round bracket
\draw[decorate,decoration={brace,amplitude=4pt}] (1.05,0.75) -- (10.3,0.75) node[midway,above=4pt,font=\scriptsize] {one round, repeated $R$ times};
% ---------- what runs ----------
\def\ys{-4.2}\def\yh{-5.55}
\node[anchor=east,font=\scriptsize\bfseries,text=blue!60!black] at (0,\ys-0.3) {simulation};
\node[anchor=east,font=\scriptsize\bfseries,text=orange!70!black] at (0,\yh-0.3) {hardware};
\draw[black!25] (-1.6,\yh+0.12) -- (12.2,\yh+0.12);
\node[note,text width=2.35cm] at (0.05,\ys) {each 1 decays with probability \mbox{$1-e^{-t/T_1}$}, the belief spreads};
\node[note,text width=3.4cm] at (2.6,\ys) {the CNOTs permute the 32 states, and the reading updates the belief by Bayes' rule};
\node[note,text width=2.3cm] at (6.25,\ys) {the record moves to the bath, heat \mbox{$=x\times$} expected 1s};
\node[note,text width=1.85cm] at (8.65,\ys) {the blamed qubit flips in the belief};
\node[note,text width=1.7cm] at (10.95,\ys) {exact error of the Bayes decision};
\node[note,text width=2.35cm] at (0.05,\yh) {the qubits really wait and decay};
\node[note,text width=3.4cm] at (2.6,\yh) {native gates, mid-circuit readout of $a_0,a_1$, no reset when the record is kept};
\node[note,text width=2.3cm] at (6.25,\yh) {three CNOTs per swap, the 1s in the bath give the heat};
\node[note,text width=1.85cm] at (8.65,\yh) {the chip waits for the reading, then applies $X$};
\node[note,text width=1.7cm] at (10.95,\yh) {thousands of shots, same decoder};
\end{tikzpicture}
\caption{One round of the code as a circuit, and what each step does in simulation and on hardware. The data qubits $d_i$ store $000$ or $111$ ($X^{\ell}$ prepares logical $\ell=1$) and wait while they decay. A check (1) copies the parities $d_0\oplus d_1$ and $d_1\oplus d_2$ onto the ancillas $a_k$ with four CNOTs and reads them. The record is then kept or swapped into a bath qubit $b_k$ (2), and the qubit blamed by the reading $s$ is corrected now (3) or later in software. Skipping a round leaves out blocks 1 to 3. The numbers match the decision points in Fig.~\ref{fig:entropy}. The inset shows how the seven qubits are connected on the chip.}
\label{fig:setup}
\end{figure*}

\subsection{Learning to act under partial information}

\emph{Reinforcement learning.} Reinforcement learning (RL) finds good decision rules by trial and error~\cite{SuttonBarto2018}. An \emph{agent} repeatedly observes the state of an environment, takes an \emph{action}, and receives a \emph{reward}. Over many episodes it adjusts its \emph{policy}, the rule mapping observations to actions, so as to maximize the total reward. No model of the environment is given to the agent, and it learns only from the consequences of its own choices. In our setting an episode is one run of the code over $R$ rounds, an action is the controller's choice in one round, and the reward is the negative of the logical error plus the heat price.

\emph{Belief states.} The controller never sees the data directly, only the syndrome readings, so it cannot know the true state of the qubits. The best it can do is maintain a probability distribution over the possible states, updated by Bayes' rule after every reading. This distribution, the \emph{belief}, summarizes everything the past readings say about the present. A decision problem in which the state is hidden but a belief can be maintained is a partially observable Markov decision process (POMDP)~\cite{Kaelbling1998}. Here the belief has only 32 entries, which is what makes an exact solution possible.

\emph{Two families of learners.} Proximal policy optimization (PPO)~\cite{Schulman2017} learns the policy directly. Picture a player who keeps a probability for each possible move and, after each batch of games, makes the moves that led to good outcomes a little more likely and the rest a little less likely. PPO limits how far these probabilities may shift in a single update, its \emph{clipped} step, so that one lucky or unlucky batch cannot undo what has been learned. A deep Q-network (DQN)~\cite{Mnih2015} works the other way round. It learns a score for every action in every situation, an estimate of the total reward that will follow if that action is taken, and then simply picks the action with the highest score. The letter Q is the conventional symbol for this score, and ``deep'' refers to the neural network that computes it. Both learners use a small neural network, a flexible function adjusted during training, to turn the belief into action probabilities or scores. A policy learner can be pulled off course by noisy rewards, while a score learner can be misled when it trusts the highest of several noisy scores. The two families therefore have different weaknesses, and Sec.~\ref{sec:learning} shows each in action.

\emph{Exploration, local optima and curricula.} Because a learner improves its policy in small steps, it can stop at a \emph{local optimum}, a policy from which every small change looks worse even though a better policy exists further away. Adding randomness to the actions (\emph{exploration}, for example through an entropy bonus in PPO) helps the learner try changes it would otherwise avoid. A \emph{curriculum} instead changes the task during training, which starts with an easy version and gradually makes it harder, so that the learner reaches the good region of policies before the hard version could trap it. A simple form is a \emph{warm start}, in which training on the harder task begins from an agent already trained on the easier one. Training under randomly varied environment parameters, called domain randomization~\cite{Tobin2017}, is a common way to make a learned policy keep working in an environment that differs from the one it was trained on.

%% ===================================================================== SETTING
\section{Setting}
\label{sec:setting}

The system is chosen so that every answer can be checked. We use the smallest code in which the keep-or-erase decision is meaningful, a noise model whose every parameter can be measured on the device, and an exact solver whose answers can be checked line by line. This lets the framework be tested where every answer is known before it is applied where none is.

\emph{Code and actions.} A logical bit is stored in three data qubits as $|000\rangle$ or $|111\rangle$, with two ancillas that collect the parities $d_0\oplus d_1$ and $d_1\oplus d_2$ (Fig.~\ref{fig:setup}). Over $R$ rounds, the controller picks one of five actions per round. It can skip, or it can check and then either keep the record on the ancillas or erase it into a bath qubit. After a check it can also apply an immediate conditional correction (``fix now'') instead of a correction applied in software at the end. The standard protocol checks, erases and fixes every round. Because an ancilla is not reset after a kept check, a second check XORs the parity in again. If the parity has not changed, the ancilla returns to $0$ and the record is \emph{uncomputed} at no heat cost. Nothing is forgotten in this step. The ancilla is returned to a known state by a reversible operation, so Landauer's bound does not apply to it (Sec.~\ref{sec:background}). At the end, whatever record remains is erased, and the data are read out and decoded.

\emph{One round in the circuit.} Figure~\ref{fig:setup} shows one round as a circuit. Followed for a stored logical $1$, it runs as follows.
\begin{enumerate}
\item \emph{Wait.} The data idle for a few microseconds, and each data qubit holding $1$ may decay to $0$, for example $111\to101$.
\item \emph{Check.} Four CNOTs leave each ancilla holding the parity of its two neighbors, and the ancillas are measured in the middle of the circuit. For $101$ both parities are $1$, so the syndrome is $11$, which blames $d_1$. The check reveals where the copies disagree but not the stored bit.
\item \emph{Keep or erase.} A kept record stays on the ancillas, which are not reset, so the next check adds the new parities to the old ones. If nothing has changed the two cancel ($p\oplus p=0$) and the ancillas return to $0$ without heat. An erasure instead swaps each ancilla into a bath qubit with three CNOTs. The ancilla is left in $0$ and every $1$ that reaches the bath carries one excitation, $x\kT$ of heat.
\item \emph{Fix now or later.} A correction applied now is a conditional $X$ on the blamed qubit, which the chip can apply only after the reading has arrived, so the data idle during this feedforward. Otherwise the correction is recorded in software and applied to the final reading.
\item \emph{Final readout.} After the last round the data and bath qubits are measured, and the Bayes decoder combines every reading into the logical value.
\end{enumerate}
In simulation the same steps act on the belief over the 32 states. Each gate permutes the states, each wait moves probability from $1$ to $0$ at the rate set by $T_1$, and each reading reweights the states by Bayes' rule. Here the logical error and the heat follow exactly, without sampling. On hardware the circuit is compiled to the chip's native gates and run for thousands of shots. Each shot returns one bit string of syndromes, data readings and bath readings, and the same decoder turns it into a logical value. Simulation and experiment thus share every step except the last, where one computes the error and the other counts it.

\emph{Noise model.} For a bit-flip code on superconducting qubits, the dominant noise is one-way energy relaxation, $1\to0$ with time constant $T_1$, together with asymmetric readout error. Our model tracks the exact classical probability distribution over the $2^5=32$ states of the data and ancillas. It includes a separate $T_1$ and readout error for every qubit, the CNOT error, the durations of gates, measurement, feedforward and reset, and a reset failure probability~\cite{Magnard2018}. Parameters come from the device calibration, from fits to dedicated hardware circuits, and, in the hardware runs, from calibration circuits embedded in the same job (Appendix~\ref{app:model}). The decoding is Bayes-optimal under the same model. In other words, the final logical value is the one with the higher posterior probability given every reading in the run, with each qubit's own decay and readout errors taken into account, rather than a plain majority vote.

\emph{Heat.} We count the excitation energy that leaves the record register, whether by a swap into the bath or by decay of an ancilla, in units of $\kT$ with $x=hf/\kT\approx7.74$ for a $5$~GHz qubit at $31$~mK, the same reference values used in Ref.~\cite{IshidaHasegawa2026}. Erasing a bit with excited-state probability $p$ into a bath at thermal population $p_b$ releases $Q/\kT=x(p-p_b)$. This exceeds the Landauer bound by $\kT D(p\Vert p_b)$, the entropy produced by erasing into a finite-temperature bath~\cite{Esposito2011,LandiPaternostro2021} (Appendix~\ref{app:heat}).

\emph{Cost.} The controller minimizes $\pe+\lambda\,Q/\kT$, the logical error plus the heat price $\lambda$ times the expected heat. Setting $\lambda=0$ asks only for protection. As $\lambda$ grows, the controller is increasingly rewarded for releasing less heat. The heat price thus sets how much the controller pays to discard entropy into the bath rather than leave it in place or return it reversibly.

\emph{Baselines.} Every controller is compared with three references. Passive majority vote, which never checks and decodes only the final readout, is the null model, since a controller that checks must beat it to justify its checks. The standard protocol, which checks, erases and corrects every round, and periodic schedules are the conventional baselines. The exact optimum bounds what any controller can achieve within the model.

%% ===================================================================== METHODS
\section{Methods}
\label{sec:methods}

The results rest on three tools, each built to answer one of the paper's three questions. We ask what the best controller is, whether an agent can learn it, and whether it holds on a real device. An exact solver finds the optimum, a training procedure lets an agent find it without being told, and a hardware protocol is designed so that the data test the model rather than fit it.

\emph{Exact solver.} Given the belief over the 32 states, every action and every possible reading can be enumerated. Each round has one skip branch and four checking actions with four possible readings each, so the decision tree has $17^R$ leaves, and the optimum is found by exhaustive backward induction. The belief is a sufficient statistic for the past, which makes the problem a fully specified partially observable Markov decision process (POMDP)~\cite{Kaelbling1998}. The procedure is summarized in Algorithm~\ref{alg:exact}, and the recursion and the belief update are written out in Appendix~\ref{app:solver}. The same recursion, with the minimum replaced by a given policy's choice, returns that policy's cost exactly.

\begin{algbox}{Exact optimization and exact evaluation. $\textsc{After}_a$ applies the rest of action $a$ (correction, erasure) to the updated belief.}\label{alg:exact}
\begin{algorithmic}[1]
\Require device model, rounds $R$, heat price $\lambda$, and optionally a policy $\pi$ to evaluate
\Function{Cost}{$t$, $b$} \Comment{$b$ is the belief}
  \If{$t=R$}
    \State \Return $\Pr[\text{error}\mid b]+\lambda\,h_{\mathrm{end}}(b)$
  \EndIf
  \State $\mathcal{A}\gets$ all actions, or $\{\pi(t,b)\}$ if $\pi$ is given
  \For{each $a\in\mathcal{A}$}
    \State $\tilde b\gets M_a\,b$ \Comment{gates and decay}
    \If{$a$ is skip}
      \State $C_a\gets \textsc{Cost}(t+1,\tilde b)$
    \Else
      \State $C_a\gets \lambda\,h(b,a)$ \Comment{heat of this action}
      \For{each reading $r\in\{00,01,10,11\}$}
        \State $P_r\gets\sum_s L_r(s)\,\tilde b(s)$
        \State $b'\gets \textsc{After}_a\!\left(L_r\odot\tilde b/P_r\right)$ \Comment{Bayes update}
        \State $C_a\gets C_a+P_r\,\textsc{Cost}(t+1,b')$
      \EndFor
    \EndIf
  \EndFor
  \State \Return $\min_{a\in\mathcal{A}} C_a$
\EndFunction
\State \Return $\textsc{Cost}(0,b_0)$ \Comment{optimum, or cost of $\pi$}
\end{algorithmic}
\end{algbox}

\emph{Reinforcement learning.} The agent observes the log-belief, the round index, the time since the last erasure and whether a bath qubit has been used. It is rewarded with the negative of the expected cost computed from its belief. We use PPO~\cite{Schulman2017} from Stable-Baselines3~\cite{Raffin2021}, with two hidden layers of 64 units, 200k training steps and 8 parallel environments. Three seeds are trained per setting at 4 rounds (two to four for the long circuits, Table~\ref{tab:rl}), and we keep the seed whose policy has the lowest cost when evaluated exactly by the solver, so that no sampling noise enters the comparison with the optimum. DQN~\cite{Mnih2015,vanHasselt2016} serves as a value-based comparison, and agents meant to withstand drift are trained with randomized device parameters~\cite{Tobin2017}. The training and selection procedure, including the curriculum in the heat price introduced in Sec.~\ref{sec:learning}, is summarized in Algorithm~\ref{alg:train}. PPO and DQN themselves are used as standard library implementations, and their settings are listed in Table~\ref{tab:rl}.

\begin{algbox}{Training with exact model selection and a heat-price curriculum. Each step is described in Sec.~\ref{sec:methods}.}\label{alg:train}
\begin{algorithmic}[1]
\Require heat prices $\lambda_1=0<\lambda_2<\dots$, seeds $k=0,1,2$, curriculum on or off
\For{each $\lambda$, in increasing order}
  \For{each seed $k$}
    \If{curriculum is on and $\lambda>0$}
      \State $\pi_k\gets\pi^\star_{0}$ \Comment{warm start}
    \Else
      \State initialize $\pi_k$ at random
    \EndIf
    \State train $\pi_k$ with PPO at price $\lambda$ (200k steps)
    \State $J_k\gets$ exact cost of $\pi_k$ \Comment{Algorithm~\ref{alg:exact}}
  \EndFor
  \State keep $\pi^\star_\lambda\gets\pi_{k^\star}$ with $k^\star=\arg\min_k J_k$
  \State report the gap $J_{k^\star}/J_{\mathrm{opt}}-1$ to the optimum
\EndFor
\end{algorithmic}
\end{algbox}

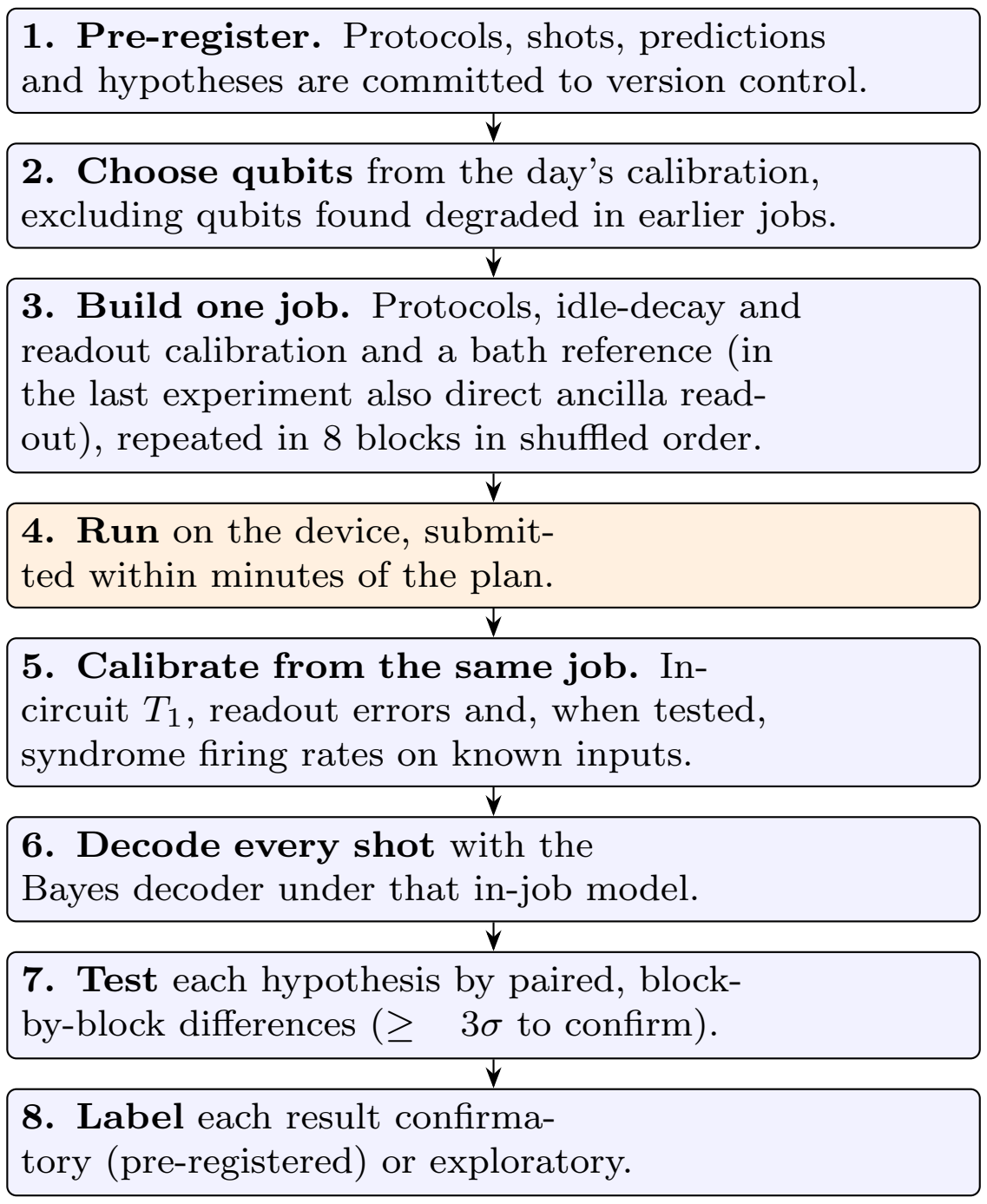
\begin{figure}[t]
\centering
\begin{tikzpicture}[font=\footnotesize, node distance=2.2mm, >={Stealth[length=1.6mm]},
  st/.style={draw, rounded corners=2pt, fill=blue!5, text width=7.2cm, align=left, inner sep=3pt},
  ob/.style={draw, rounded corners=2pt, fill=orange!12, text width=7.2cm, align=left, inner sep=3pt}]
\node[st] (p) {\textbf{1. Pre-register.} Protocols, shots, predictions and hypotheses are committed to version control.};
\node[st, below=of p] (q) {\textbf{2. Choose qubits} from the day's calibration, excluding qubits found degraded in earlier jobs.};
\node[st, below=of q] (j) {\textbf{3. Build one job.} Protocols, idle-decay and readout calibration and a bath reference (in the last experiment also direct ancilla readout), repeated in 8 blocks in shuffled order.};
\node[ob, below=of j] (r) {\textbf{4. Run} on the device, submitted within minutes of the plan.};
\node[st, below=of r] (c) {\textbf{5. Calibrate from the same job.} In-circuit $T_1$, readout errors and, when tested, syndrome firing rates on known inputs.};
\node[st, below=of c] (d) {\textbf{6. Decode every shot} with the Bayes decoder under that in-job model.};
\node[st, below=of d] (t) {\textbf{7. Test} each hypothesis by paired, block-by-block differences ($\ge3\sigma$ to confirm).};
\node[st, below=of t] (l) {\textbf{8. Label} each result confirmatory (pre-registered) or exploratory.};
\foreach \a/\b in {p/q,q/j,j/r,r/c,c/d,d/t,t/l} \draw[->] (\a) -- (\b);
\end{tikzpicture}
\caption{Protocol followed by every hardware experiment after the first. Steps 1--3 are fixed before submission, and steps 5--7 use only data from the same job.}
\label{fig:protocol}
\end{figure}

\emph{Hardware protocol.} Experiments ran on \texttt{ibm\_marrakesh} and \texttt{ibm\_kingston}, two 156-qubit Heron processors, through Qiskit~\cite{Qiskit2024}. We also use dynamic circuits~\cite{Corcoles2021}, which are circuits whose later operations depend on earlier measurement results. Each circuit is run many times, and each run is one \emph{shot}. From the second experiment on, each job interleaves the protocols with calibration circuits (idle decay over up to 27~$\mu$s, bare readout, and a bath reference) in 8 \emph{blocks} (6 in the job that measured the reset failure), each a pass through every circuit in a freshly shuffled order. Slow drift therefore affects all protocols alike, and comparisons are \emph{drift-paired}, meaning that differences are taken within each block and then averaged. Decoding uses only the calibration measured in the same job, including the in-circuit $T_1$, which is the decay time measured inside the job itself rather than taken from the published calibration. For every confirmatory run, the protocols, shot counts, predictions and hypotheses were committed to version control before submission, a practice known as \emph{pre-registration}~\cite{Nosek2018}. Analyses that were not fixed in advance are labeled exploratory. Figure~\ref{fig:protocol} summarizes the protocol.

%% ===================================================================== OPTIMUM
\section{What the optimal controller does}
\label{sec:optimum}

What does the best controller do with its freedom? The short answer is that it checks once and erases almost nothing. We first describe the exact optimum at 4 rounds across all heat prices, then explain the three features that give it its form, and finally show that the same timetable wins in longer circuits.

\subsection{The optimum checks once, at the end}

\begin{figure*}[t]
\centering
\includegraphics[width=\textwidth]{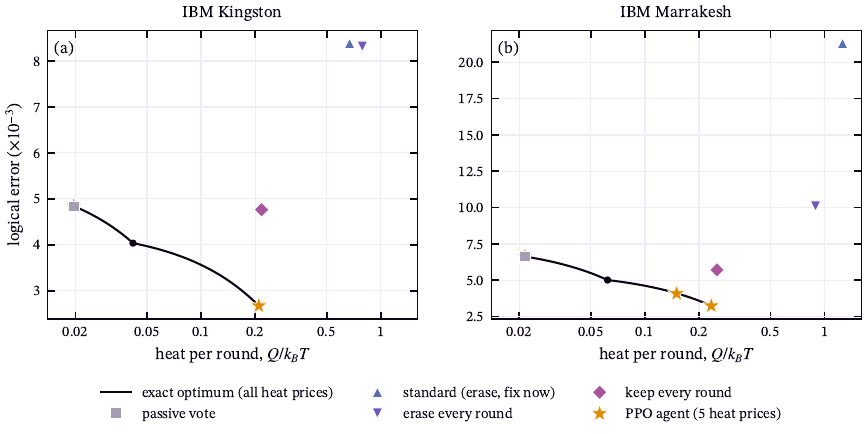}
\caption{No fixed protocol reaches the optimal error--heat frontier, the lowest error achievable for each amount of heat. Shown at 4 rounds for the (a) Kingston and (b) Marrakesh models, with the exact optimum (black), PPO at five heat prices (stars) and the fixed protocols. Segments between optimal policies are reached by mixing them.}
\label{fig:frontier}
\end{figure*}

Figure~\ref{fig:frontier} shows the exact error--heat frontier at $R=4$. At $\lambda=0$ the optimum skips three rounds and checks once at the end, then keeps the record. Its logical error is $2.7\times10^{-3}$ on the Kingston model, against $4.8\times10^{-3}$ for passive majority vote and $8.4\times10^{-3}$ for the standard protocol. On the Marrakesh model the same three numbers are $3.2\times10^{-3}$, $6.6\times10^{-3}$ and $21.3\times10^{-3}$, the last inflated by that model's longer decision wait for real-time correction (Appendix~\ref{app:model}). At intermediate prices (from $\lambda\approx0.002$ on Kingston and $0.003$ on Marrakesh up to $\lambda\approx0.01$, shown for $\lambda=0.003$) the optimum checks at round 3 and checks again at round 4 only if the first check fired. This undoes the record instead of erasing it and cuts the heat per round from $0.21$ to $0.04\,\kT$. Above $\lambda\approx0.01$ the optimum never checks.

Three features of the problem explain why the optimum takes this form, and each can be traced to the physics of the device. The first is \emph{skipping}. On these devices a check adds gate errors, readout errors and dead time faster than it removes errors, so a check pays only when its information is used soon after. The second is \emph{keeping instead of erasing}. A kept record can be uncomputed for free ($p\oplus p=0$), while erasing it costs heat and, through reset failure, also error. Simulations of surface-code memories already show that skipping the ancilla reset costs nothing in logical performance~\cite{Geher2025}, and the heat accounting here adds a thermodynamic reason to prefer it. The third is \emph{postponing correction}. Flipping a decayed qubit back to $1$ re-exposes it to decay for the rest of the circuit, whereas a correction applied in software at readout has no such cost. Real-time correction never appears in the optimum at any heat price. In terms of entropy flow, the optimum lets the entropy of the data build up during the idle rounds and removes it once, at the point where the information is used, and when heat is priced it returns the record reversibly rather than discarding it.

\subsection{The same timetable wins in longer circuits}

\begin{figure*}[t]
\centering
\includegraphics[width=\textwidth]{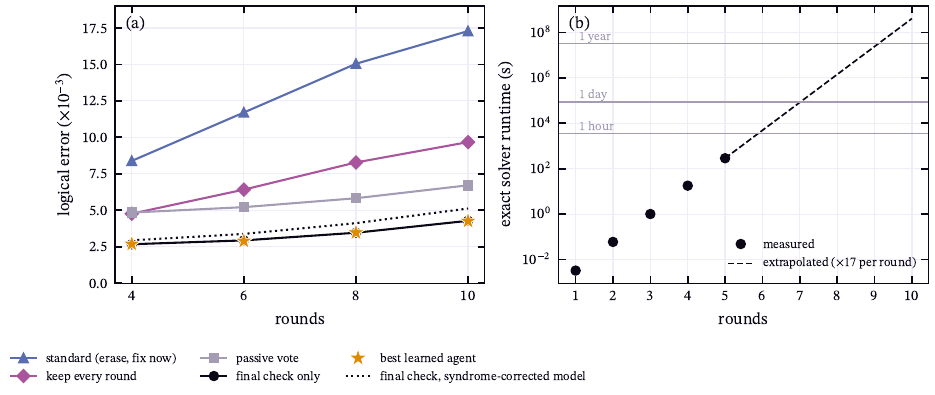}
\caption{A single check before readout beats every fixed schedule at every length. (a) Logical error against circuit length at $\lambda=0$ (Kingston model) for the fixed protocols and the best learned agent (stars). The dotted line shows the final check with the syndrome correction of Sec.~\ref{sec:hardware}. (b) The exact solver's runtime grows by a factor of 17 per round.}
\label{fig:long}
\end{figure*}

This result is not an artifact of a short circuit. We compared fixed timetables from 4 to 10 rounds [Fig.~\ref{fig:long}(a)]. A single check before readout gives $\pe=2.7$, $2.9$, $3.5$ and $4.3\times10^{-3}$ at 4, 6, 8 and 10 rounds. This is between 36\% and 44\% below the best of every other fixed schedule we tried, namely passive vote, keeping or erasing every round, the standard protocol, and periodic checking every 2, 3 or 4 rounds. Its error grows with length only through decay during the idle rounds.

%% ===================================================================== LEARNING
\begin{figure*}[t]
\centering
\includegraphics[width=\textwidth]{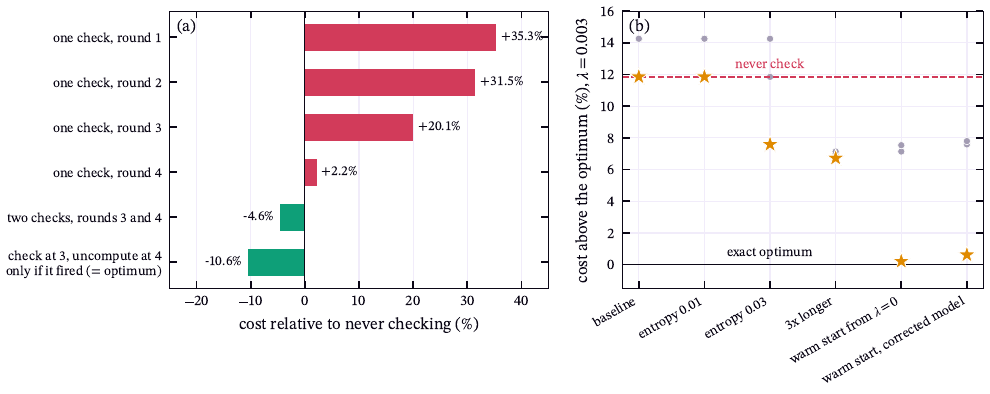}
\caption{Gradient-based learning stalls at ``never check'' because no single change improves on it ($\lambda=0.003$, 4 rounds, Kingston model). (a) Exact cost of policies one or two changes away. Every single added check is worse, but two checks together can be better. (b) PPO's cost above the optimum under different training settings (dots are seeds, stars the best of three). Only a warm start from the $\lambda=0$ agent reaches the optimum.}
\label{fig:trap}
\end{figure*}

\section{Learning the controller}
\label{sec:learning}

A learning agent that is never shown the optimum recovers it in almost every setting we tested, and the few settings where it falls short reveal something about the problem itself. The exact optimum makes this comparison possible, since it gives a benchmark that is rarely available in learned quantum control. We first verify that PPO recovers the optimum where brute force stops, and then examine the two places where it falls short.
\subsection{PPO recovers the optimum where brute force stops}

At 4 rounds the PPO agent reproduces the exact optimum, with identical behavior, at four of the five heat prices on both device models. At $\lambda=0.003$ it is $11.9\%$ (Kingston) and $2.2\%$ (Marrakesh) above it. At 4--10 rounds the learned agents find the single-late-check timetable at every length and match it to within $\pm1.6\%$ [Fig.~\ref{fig:long}(a), stars]. At 10 rounds, an adaptive agent that could in principle condition on every reading does no better than the single late check (within $0.6\%$). We can see that adaptivity adds nothing here, and this is also a statement about model complexity. At $\lambda=0$ an agent that sees only the round number, the simplest policy class we tried, finds the same timetable as agents that see the full belief at 4, 6 and 8 rounds. At 10 rounds it settled on two final checks instead, $11\%$ worse than the single late check.

Learning is needed because the exact solver's runtime grows as $17^R$ [Fig.~\ref{fig:long}(b)]. In our single-threaded implementation it took about 18~s at 4 rounds and about 5~min at 5. Each further round multiplies the cost by about 17, so the same search would take hours at 6 rounds, weeks at 8 and years at 10. Training the RL agents took less than an hour at every length we tested. A smarter exact method, for example one that merges identical beliefs or prunes dominated branches, could reach further. The point is that brute force stops near 6 rounds, while the learner needs no problem-specific shortcuts built by hand.

\subsection{What makes learning hard}

\emph{The ``never check'' trap.} The one heat price at which PPO missed the optimum turns out to be the most informative. Figure~\ref{fig:trap}(a) gives the exact cost of every policy one step away from ``never check''. Adding a single check at any round raises the cost, by $+2.2\%$ to $+35.3\%$, and only two coordinated changes help. Two fixed checks at rounds 3 and 4 lower the cost by $4.6\%$, and the optimum, which makes the second check conditional on the first, lowers it by $10.6\%$. ``Never check'' is therefore a genuine local optimum of the problem, a barrier two actions wide, and a learner that improves one decision at a time sees every step lead uphill. A gradient-based learner such as PPO moves its policy in small steps toward lower cost. When every small step raises the cost, it receives no signal pointing toward the better policy two steps away, however long it trains.

\begin{figure*}[t]
\centering
\includegraphics[width=\textwidth]{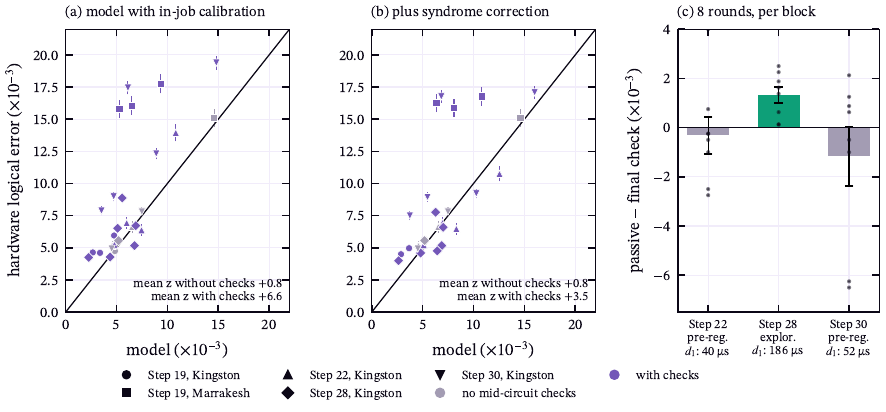}
\caption{Protocols with checks fare worse on hardware than the model predicts. (a) Logical error of every protocol in five jobs against the model with in-job calibration. Protocols without checks (light) lie on the diagonal, and those with checks (dark) lie above it. (b) The same after the syndrome correction, fitted on known-input syndrome rates only. (c) Late check versus passive vote at 8 rounds, block by block, against the in-job $T_1$ of the middle data qubit.}
\label{fig:hardware}
\end{figure*}

Figure~\ref{fig:trap}(b) shows the consequence. With the original settings PPO never checks ($+11.9\%$ above the optimum). More exploration (an entropy bonus of $0.03$) or three times longer training gets over the barrier to the fixed pair ($+7.6\%$ and $+6.7\%$) but not to the conditional optimum. Starting instead from the $\lambda=0$ agent, which already checks, and then raising the price reaches the optimum ($+0.2\%$). The same curriculum works on the syndrome-corrected model of Sec.~\ref{sec:hardware} ($+0.6\%$). The intuition is that the useful behavior must first be learned where it is cheap, and this lesson reaches beyond error correction. Whenever a benefit appears only if a later action is conditioned on an earlier one, gradient-based learners settle on the trivial policy, and a curriculum in the cost parameter that starts where the useful behavior is cheap is a simple remedy.

\emph{Value overestimation in DQN.} DQN is mixed at low heat prices. It is exact on the Marrakesh model at $\lambda=0$ and $0.001$, but $+25.5\%$ above the optimum at $\lambda=0$ on Kingston, although at $\lambda=0.003$ it does better than PPO there ($+7.5\%$). At high prices it fails badly, at $+367\%$ and $+936\%$ above the optimum at $\lambda=0.01$ and $0.03$ on Kingston, and $+35\%$ and $+720\%$ on Marrakesh. Comparing its learned values with the exact ones at the first decision shows why. They are compressed and optimistic, and on Kingston they rank erasing first where erasing is among the costliest actions. This is the maximization bias that double Q-learning was introduced to reduce~\cite{vanHasselt2016}. The intuition is that DQN picks the action with the best estimated value, and when the estimates are noisy, the action that looks best is disproportionately often one whose value happens to be overestimated.

\emph{Drift and domain randomization.} Device parameters also drift between calibrations~\cite{Klimov2018}. At $\lambda=0$ every policy we tested stays within $+3.8\%$ (Kingston) and $+6.1\%$ (Marrakesh) of the drift-aware oracle, the optimum computed for the true, drifted device, in the worst case over the drift range. At $\lambda=0.003$, however, every fixed policy becomes brittle, with a worst case of $+32.8\%$ for the nominal optimum on Kingston, and training under randomized parameters did not help ($+46.9\%$, the same as the nominal agent). Domain randomization asks the agent to do well on average over many plausible devices. Its worst case here equals that of the nominal agent, which never checks, consistent with the randomized training settling on the same policy that is safe on every device and that the trap described above also favors. The practical implication is that, at a nonzero heat price, holding up under drift comes from calibrating inside the job, not from a more conservative fixed policy.

%% ===================================================================== HARDWARE

\section{Hardware}
\label{sec:hardware}

On a real device the simulated advantage survives only in part. Frequent checking is confirmed to cost logical error, but the single late check wins only on healthy qubits, and finding out why took several experiments. We first report a failed pre-registered prediction and the short-circuit experiments that followed, then trace the model's miss to the syndromes, and finally return to whether the late check beats passive vote on a real device.

\subsection{Frequent checking costs error, and the model misses only the checks}

\emph{A failed prediction.} Our first pre-registered experiment (\texttt{ibm\_marrakesh}, 40,000 shots) missed badly. Passive vote, predicted at $7.0\times10^{-3}$, measured $23.6\times10^{-3}$, and the learned protocols lost their predicted advantage. The diagnosis identified three causes. One qubit's readout error was $2.1\%$ rather than the calibrated $0.1\%$, the decoder trusted those calibrated numbers, and $T_1$ drifted between calibration and run. Two parameters that we had fitted indirectly were also wrong. Measured directly in a later job on \texttt{ibm\_kingston}, the decision wait for real-time correction was 0.05--0.23~$\mu$s rather than the fitted 10~$\mu$s, and the reset failure was 1.2--3.5\%. All later runs therefore embed their own calibration. We report this failure transparently because it shaped every experiment that followed.

\emph{Learned protocols beat frequent checking.} With embedded calibration on \texttt{ibm\_kingston} (4,000 shots $\times$ 8 blocks), both learned protocols beat checking every round, by $0.0013\pm0.0004$ ($3.6\sigma$) and $0.0013\pm0.0003$ ($4.4\sigma$), in 7 of 8 blocks each. The intermediate-price protocol did so with $38\%$ less heat ($0.157$ versus $0.253\,\kT$ per round). \texttt{ibm\_marrakesh} pointed the same way ($1.6\sigma$ and $2.0\sigma$). Passive vote, however, tied the single late check at 4 rounds ($+0.0001\pm0.0005$), whereas the model had predicted a $45\%$ gap.

\emph{The model fails only for checked protocols.} Figure~\ref{fig:hardware}(a) compares the measured and predicted logical error of every protocol in five jobs, and the pattern is sharp. We measure the mismatch by $z$, the difference between measurement and prediction in units of its uncertainty, so that $z\approx0$ means the model is right. Protocols without mid-circuit checks are predicted well in every job (mean $z=+0.8$), while protocols with checks come out clearly worse than predicted (mean $z=+6.6$). The missing cost must therefore lie in the checks themselves, and to identify it we ran three further experiments on \texttt{ibm\_kingston}.

\subsection{Checks do not damage the data, but syndromes misfire}

\emph{No damage to the data.} A pre-registered experiment measured the raw error of the data qubits, without decoding, after 0, 1, 2, 4 or 8 checks whose results are kept but never used, and after a single check placed early, in the middle or last. Because such checks never feed back, the data evolve in the same way whatever the readings, so the raw error isolates what a check does to the data. The extra raw error over the checked schedules was $+0.0001\pm0.0002$, that is, none. If anything the model overestimates the decay during check rounds. With 8 checks, $0.0607$ was measured against $0.0681$ predicted. An earlier and simpler correction, one extra data flip per check fitted on other jobs, had explained the tie between the late check and passive vote but overshot protocols with many checks. This measurement shows that the extra cost is not damage to the data.

\emph{Syndromes are less reliable during checks.} The data start in $000$ or $111$, so a syndrome bit should fire only through an error. Its firing rate on known inputs therefore measures syndrome reliability without using the logical outcomes. In every job, on both devices, the second ancilla raised false alarms far more often than the model predicts. Fitted as an extra readout-like error, the excess was $0.025$--$0.040$ on Kingston and $0.034$ on Marrakesh, while the first ancilla needed little or none on Kingston ($\le0.004$). This held on different physical qubits, which points at the ancilla's role in the circuit rather than at one bad qubit. Read out directly, without a check, the same ancilla had readout errors of only $0.4\%$ and $1.0\%$. The false alarms therefore arise during the check, from the entangling gates or an excitation they cause, and not in readout.

In a pre-registered replication, this correction lowered the misfit of the checked protocols ($\chi^2$ from 897 to 603), and over all jobs it halves the mean deviation [Fig.~\ref{fig:hardware}(b), mean $z=+3.5$]. Re-optimizing on the corrected model leaves the optimal controller unchanged at every heat price. Only the size of the late check's advantage shrinks [Fig.~\ref{fig:long}(a), dotted], and PPO again matches the corrected optimum.

\subsection{The late check on hardware}

The question that remains is whether the late check beats passive vote on a real device. Three jobs compared them at 8 rounds [Fig.~\ref{fig:hardware}(c)]. In the first pre-registered run, the middle data qubit had degraded to $T_1\approx40\,\mu$s, well below the $282\,\mu$s in the device calibration, and the two protocols tied ($-0.0003\pm0.0007$). In the check-cost experiment, on healthy qubits ($T_1=174$--$191\,\mu$s), the late check won by $0.0013\pm0.0003$ ($4.0\sigma$, 8 of 8 blocks, an exploratory comparison). This 24\% reduction is comparable to the 29\% that the corrected model predicts for the reference device parameters. In the pre-registered confirmation an hour later, the same middle qubit had dropped to $T_1\approx50\,\mu$s and the advantage vanished ($-0.0012\pm0.0012$, not confirmed). The middle qubit's $T_1$ was stable within each job, so these data cannot separate its condition from other differences between jobs. We therefore report the link between the middle qubit's health and the late check's advantage as an association, and not as a demonstrated cause. In all three runs, however, checking every round stayed clearly worse than the late check, and in the pre-registered confirmation both checking every round and correcting in real time lost to it, by $3.6\sigma$ and $10.1\sigma$.

%% ===================================================================== DISCUSSION
\section{Discussion}
\label{sec:discussion}

We discuss the results in three parts, covering what they mean for error correction, what they mean for learned quantum control and what the hardware adds, and finally the limitations, the extensions they point to and the practical recommendations that follow.

\subsection{Error correction as a controlled flow of entropy}

Seen thermodynamically, each round of error correction moves entropy out of the data. Errors raise the entropy of the data qubits, a check copies that uncertainty into the ancillas, and an erasure discards it into the bath as heat. In this sense a QEC cycle works as a refrigerator for the logical information~\cite{Landi2020QECengine,Czartowski2026} (Fig.~\ref{fig:entropy}). The protocol assumed by most current experiments runs this refrigerator at full duty. Every round is checked, every record is erased, and every fired syndrome is corrected at once. For a small code under realistic noise, the results presented here show that the controller that best protects the logical bit runs it almost never. It removes the entropy once, just before readout, and when heat is priced it returns a kept record to its initial state by uncomputation, which releases no heat, instead of discarding it.

This sharpens recent evidence that checking less often can help~\cite{AbuNada2017,MartonAsboth2025,Haug2026,Berthusen2025,SohnOh2026} in two ways. First, the optimum does not merely check less. It keeps the record and undoes it, so the heat it releases is a small fraction of the standard protocol's. Second, the same timetable is optimal when heat costs nothing, so in this regime protection and low dissipation do not trade off. Our approach is also complementary to making each erasure cheaper~\cite{IshidaHasegawa2026}, since a controller that erases rarely could erase into a recycled bath.

The broader point is that the cycle of checks, erasures and corrections need not be fixed in advance. Once it is posed as a control problem with a heat price, it becomes a design variable. A controller can decide how much entropy to remove, when to remove it, and whether to discard it or return it reversibly. The action set and the cost are defined for any stabilizer code. What grows with the code is the size of the belief and of the decision tree, which is where learning becomes necessary.

\subsection{Learned control, checked exactly and on hardware}

A recurring difficulty in learned quantum control is that there is rarely a known answer to check the learner against. Here the exact solver supplied one, and it turned RL from a black box into something we could test. The agent recovered the optimum wherever it is known, and its one failure traced to a feature of the problem, a local optimum that no single change escapes, which has a simple remedy. The value overestimation of DQN and the failure of domain randomization at a nonzero heat price are likewise properties that anyone training such a controller can anticipate. The ``never check'' trap arises from the formulation of the cost rather than from this particular code, so it is worth checking for whenever a learned controller is trained with a price on a resource. Because the learner has been validated against the exact optimum at the sizes where that optimum is available, its answers at larger sizes rest on more than the learner's own training curve.

The learned policies are also fully interpretable. Each can be read off as a short decision tree and compared action by action with the exact optimum, and the exact action values explain the failure of DQN as maximization bias rather than leaving it as an empirical observation. Physical knowledge entered the learner in four places, namely the Bayes belief as its observation, the logarithmic encoding that makes rare alarms visible, the exact solver used for model selection, and the heat-price curriculum. In this problem, learning and physical insight worked best together, and neither alone would have produced a controller that is both optimal and understood.

The hardware results show where an advantage found in simulation is lost on a real device, and they do so with a method that can be reused. The model is exact for protocols without checks, checks do not damage the data, and the loss lies in syndromes that are less reliable during checks than calibration suggests. Mid-circuit measurement is known to carry errors beyond those of terminal readout~\cite{Rudinger2022,Hothem2025,Sank2016}. Our measurement isolates its effect on the syndrome from its effect on the data, using a few extra circuits in the same job and no logical outcomes. The same in-job calibration applies unchanged to any code whose data can be prepared in known states. What remains open is why the late check's advantage disappeared in the two runs where a data qubit had degraded, even though the model, given that qubit's measured $T_1$, still predicted it.

\subsection{Limitations, next steps and recommendations}

The code is a three-qubit bit-flip code under energy relaxation, so phase errors are neither protected against nor modeled. The conclusions therefore apply to bit-flip protection only. Our heat accounting counts excitations leaving the record register. The heat measured in one 8-round job came out 1.7 to 3.5 times the pre-registered prediction for every protocol, and we do not use it for any claim. The heat axis of the frontier is therefore supported by the model, and not by a hardware measurement. The exact solver is brute force and stops near 6 rounds, which limits the range over which the learner can be checked exactly. Finally, the hardware statistics rest on a handful of jobs on two devices, and some comparisons, including the $4.0\sigma$ advantage of the late check on healthy qubits, are exploratory. Whether that advantage survives replication is open. We also did not vary the network size or the training budget systematically, beyond the round-number agent and the longer training of Sec.~\ref{sec:learning}, so how the learner scales with model and data size remains to be characterized. These limits mark out where the conclusions apply, but within that range they do not weaken them.

Three recommendations follow from the experimental findings. First, schedule checks rather than run them every round. In our model of current devices a check pays only when its information is used soon after, and in simulation a single check before readout outperforms every periodic schedule we tested. Second, calibrate the syndromes as well as the data inside each job. Syndrome firing rates on known inputs cost a few extra circuits, use no logical outcomes, and reveal errors that terminal-readout calibration misses. Third, when training a learned controller with a cost on a resource such as heat, start where that cost is zero and raise it gradually. This curriculum led out of the ``never check'' trap on both models we tested (best of three seeds). These are tractable additions that require no change to the underlying hardware.

%% ===================================================================== CONCLUSION
\section{Conclusion}

We have shown that, for a three-qubit repetition code on current superconducting hardware, the controller that best protects a logical bit erases almost nothing. It checks once before readout, and when heat is priced it undoes its record rather than erasing it. The single late check lowers the logical error by 36--44\% relative to every fixed schedule we tested in simulation. An exact solver and a PPO agent reach this answer independently, and the agent continues to find it where brute-force search becomes impractical.

On the learning side, the most revealing finding is the ``never check'' trap. The problem has a local optimum that no single change escapes, gradient-based learners settle on it, and a curriculum in the heat price leads them out. On the hardware side, the most informative finding is where the simulated advantage goes. It is not lost to damage of the data, which we measured to be absent, but to syndromes that misfire during checks and to a sensitivity to degrading qubits that remains open.

The code studied here is deliberately small so every claim about the optimum could therefore be checked exactly, and every deviation on hardware could be traced. Within that scope, the work provides three things that carry over to larger codes. The first is a formulation of error correction as the control of entropy flow out of the data, with heat as an explicit cost. The second is a way of validating a learned controller against an exact optimum before relying on it where no optimum is available. The third is an in-job calibration that locates the gap between simulated and real performance. The finding that much of the erasure in standard protocols buys no protection is established for this code on current devices. Whether it holds more generally is a question this framework can now address.

\begin{acknowledgments}
We acknowledge the use of IBM Quantum services for this work. The views expressed are those of the author and do not reflect the official policy or position of IBM or the IBM Quantum team.
\end{acknowledgments}

%% ===================================================================== APPENDICES
\appendix

\section{Noise model and parameters}
\label{app:model}

The model evolves the probability vector over the 32 basis states of $d_0d_1d_2a_0a_1$. Energy relaxation acts on every qubit during every interval with its own in-circuit $T_1$. Readout of qubit $i$ returns the wrong value with probability $r^{(i)}_{01}$ or $r^{(i)}_{10}$ depending on its state, and each CNOT fails with probability $e_{\mathrm{cx}}$. Erasing swaps an ancilla into a bath qubit that is reset with failure probability $\varepsilon$, and real-time correction waits $t_{\mathrm{ff}}$. Table~\ref{tab:params} lists the Kingston values used for the simulations of Secs.~\ref{sec:optimum} and~\ref{sec:learning}. The Marrakesh model (data qubits 92, 98 and 110) was fixed before the direct measurements of Sec.~\ref{sec:hardware} and keeps the earlier indirect values $t_{\mathrm{ff}}=10\,\mu$s and $\varepsilon=0.015$. Its fix-now protocols, including the standard protocol, are therefore penalized more than they would be with the measured decision wait, which does not affect the optimum, since the optimum never corrects in real time. Hardware analyses replace $T_1$ and the data readout errors with values measured inside each job.

\begin{table}[ht]
\caption{Kingston model (data qubits 57, 68 and 70, ancillas 67 and 69, baths 66 and 78).}
\label{tab:params}
\begin{ruledtabular}
\begin{tabular}{lc}
quantity & value \\ \hline
data $T_1$ in circuit ($\mu$s) & 79, 192, 232 \\
data readout $r_{10}$ & 0.028, 0.008, 0.017 \\
ancilla readout $r_{01}$, $r_{10}$ & 0.010/0.016, 0.002/0.011 \\
CNOT error $e_{\mathrm{cx}}$ & 0.0025 \\
CNOT, measurement, reset (ns) & 132, 2180, 2212 \\
passive round (ns) & 2452 \\
decision wait $t_{\mathrm{ff}}$ (ns, measured) & 54 \\
reset failure $\varepsilon$ (measured) & 0.023 \\
\end{tabular}
\end{ruledtabular}
\end{table}

\begin{table}[ht]
\caption{Learning settings (Stable-Baselines3~\cite{Raffin2021}). Unlisted settings are the library defaults. Long circuits use $5\times10^4$ training steps per round and two seeds, raised to $10^5$ steps per round and four seeds at 8 and 10 rounds. The DQN exploration rate decays linearly from 1 to 0.02 over the first 30\% of training.}
\label{tab:rl}
\begin{ruledtabular}
\begin{tabular}{lcc}
setting & PPO & DQN \\ \hline
hidden layers & 64, 64 & 64, 64 \\
training steps & $2\times10^5$ & $2\times10^5$ \\
parallel environments & 8 & 1 \\
learning rate & $3\times10^{-4}$ & $1\times10^{-4}$ \\
discount $\gamma$ & 1 & 1 \\
rollout per environment & 256 & -- \\
minibatch size & 256 & 256 \\
epochs per update & 10 & -- \\
GAE parameter & 1 & -- \\
entropy bonus & 0.003 & -- \\
replay buffer & -- & $2\times10^5$ \\
learning starts & -- & 5000 \\
train frequency (steps) & -- & 4 \\
target update (steps) & -- & 500 \\
exploration $\epsilon$ & -- & $1\to0.02$ \\
seeds per setting & 3 & 3 \\
\end{tabular}
\end{ruledtabular}
\end{table}

\section{Exact solver and belief update}
\label{app:solver}

Let $b_t$ be the belief, a probability vector over the 32 states of $d_0d_1d_2a_0a_1$, after $t$ rounds. An action $a$ first applies its gates and the decay during them, $\tilde b = M_a b_t$, where $M_a$ is a stochastic matrix built from the CNOT error and the $T_1$ of each qubit. If $a$ checks, the ancilla reading $r\in\{00,01,10,11\}$ occurs with probability $P(r\mid b_t,a)=\sum_s L_r(s)\,\tilde b(s)$, where $L_r(s)$ is the probability of reading $r$ in state $s$ given the ancilla readout errors. The belief is then updated by Bayes' rule,
\begin{equation}
b_{t+1}(s)=\frac{L_r(s)\,\tilde b(s)}{P(r\mid b_t,a)},
\end{equation}
followed by the decay during the measurement and, for an erasing action, the swap into the bath. In words, states that are consistent with the reading gain weight and the others lose it. The exact optimum is the solution of the backward recursion
\begin{equation}
V_t(b)=\min_a\Big[\lambda\,h(b,a)+\sum_r P(r\mid b,a)\,V_{t+1}\big(b'(b,a,r)\big)\Big],
\end{equation}
where $h(b,a)$ is the expected heat released by action $a$ and $b'$ is the updated belief. At $t=R$, $V_R(b)$ is the probability of a wrong Bayes decision on the final readout plus $\lambda$ times the heat of the final erasure. A skip has a single successor and a check has four, so the recursion visits $17^R$ leaves. Any fixed or learned policy is evaluated exactly by the same recursion with the minimum replaced by that policy's choice.

\section{Heat and the Landauer bound}
\label{app:heat}

Consider a record bit whose excited state has energy $hf$ and probability $p$. Erasing it brings it to the thermal population $p_b=e^{-x}/(1+e^{-x})$ of the bath, with $x=hf/\kT$, and releases the heat $Q=hf\,(p-p_b)$. In units of $\kT$ this heat splits exactly into two parts,
\begin{equation}
\frac{Q}{\kT}=x\,(p-p_b)=\big[H(p)-H(p_b)\big]+D(p\Vert p_b),
\end{equation}
where $H$ is the binary entropy in nats and $D$ is the relative entropy. The first term is the Landauer minimum, the entropy removed from the record, which equals $\ln2$ for a fully random bit erased at low temperature. The second term is the excess dissipated by erasing into a bath at finite temperature~\cite{Esposito2011}. For a random bit ($p=1/2$) at $x\approx7.74$, the heat is $3.87\,\kT$, of which $0.69\,\kT$ is the Landauer minimum and $3.18\,\kT$ is excess. A record that is uncomputed rather than erased is returned to $0$ by a reversible operation and releases no heat in either term.

\section{Hardware jobs}
\label{app:jobs}

Table~\ref{tab:jobs} lists every IBM Quantum job used in this work, so that each run can be retrieved and checked.

\onecolumngrid
\vspace{8pt}
\begin{center}
\refstepcounter{table}\label{tab:jobs}
\small TABLE~\thetable. IBM Quantum jobs used in this work. The design column gives shots per circuit times the number of shuffled blocks.\par\vspace{4pt}
\begin{ruledtabular}
\begin{tabular}{llll}
purpose & device & job & design \\ \hline
smoke test & Marrakesh & \texttt{dbt67tnq0h0s73bkcve0} & -- \\
pre-registered prediction & Marrakesh & \texttt{dbt9mg9vvpic738mn7t0} & 40k shots \\
in-situ calibration (diagnosis) & Marrakesh & \texttt{dbtc74fq0h0s73bkjdbg} & -- \\
short circuits, embedded cal. & Kingston & \texttt{dbtcjpnasvbs73b9j0dg} & $4000\times8$ \\
short circuits, embedded cal. & Marrakesh & \texttt{dbtcjn9vvpic738mqdg0} & $2000\times8$ \\
reset failure, decision wait & Kingston & \texttt{dbtgc23eq3sc73evpijg} & -- \\
long circuits (reduced) & Kingston & \texttt{dbtp9vmasacs73fmsgdg} & $2000\times8$ \\
check cost & Kingston & \texttt{dbv80r5arooc73fl31a0} & $4000\times8$ \\
confirmation & Kingston & \texttt{dbv8tcdarooc73fl4f1g} & $4000\times8$ \\
\end{tabular}
\end{ruledtabular}
\end{center}
\vspace{8pt}
\clearpage
\twocolumngrid

\bibliography{refs}

\end{document}